\documentclass[11pt]{article}
\usepackage{cite}
\usepackage{amsmath}
\usepackage{epsf}
\usepackage{graphics}
\usepackage{rotating}
\usepackage[small]{caption}

\usepackage{graphics}
\usepackage{amssymb}
\usepackage{subfigure}
\usepackage{epsf}
\usepackage{rotating}
\usepackage{pstricks}
\def\ba{\begin{eqnarray}}
\def\ea{\end{eqnarray}}

\def\ds#1{#1\kern-1ex\hbox{/}}
\def\dsh{h\kern-1.2ex /}

\newcommand{\bea}{\begin{eqnarray}}
\newcommand{\eea}{\end{eqnarray}}

\def\nn{\nonumber}
\def\beq{\begin{equation}}
\def\eeq{\end{equation}}

\def\beqn{\begin{eqnarray}}
\def\eeqn{\end{eqnarray}}
\def\ba{\begin{eqnarray}}
\def\ea{\end{eqnarray}}

\newcommand{\beqa}{\begin{eqnarray}}
\newcommand{\eeqa}{\end{eqnarray}}

\begin{document}

\begin{center}
\vspace{1.5cm}
{\bf \Large  An Axion-Like Particle from an $SO(10)$ Seesaw with $U(1)_{X}$}

\vspace{1.5cm}

{\bf Claudio Corian\`o, Paul H. Frampton,\\}
\vspace{0.5cm}
{\bf Alessandro Tatullo and Dimosthenis Theofilopoulos\\}
\vspace{1cm}
{\it  Dipartimento di Fisica, Universit\`{a} del Salento \\
and  INFN Sezione di Lecce,  Via Arnesano 73100 Lecce, Italy\\}

\begin{abstract}
We investigate the decoupling of heavy right handed neutrinos in the context of an SO(10) GUT model, where a remnant anomalous symmetry is $U(1)_{X}$. In this model the see-saw mechanism which
generates the neutrino masses is intertwined with the Stueckelberg mechanism, which leaves the CP-odd phase of a very heavy Higgs in the low energy spectrum as an axion-like particle. Such pseudoscalar is predicted to be ultralight, in the $10^{-20}$ eV mass range.
In this scenario, the remnant anomalous $X$ symmetry of the particles of the Standard Model is interpreted as due to the incomplete decoupling of the right handed neutrino sector. We illustrate this scenario including its realisation in the context of SO(10). 
\end{abstract}
\end{center}

\newpage
\section{Introduction} 

\noindent

Recently there has been considerable interest in the occurrence of axion-like particles
\cite{Ringwald1,Ringwald2,Ringwald3} including the appearance in model
building af anomalous $U(1)$ symmetries with a Stueckelberg field \cite{Coriano1,Coriano2,Coriano3,Coriano4,Coriano5,Coriano6,Coriano7,Coriano8,Coriano9,Coriano10,Coriano11,Coriano12,Coriano13,Coriano14,Coriano:2018uip,Coriano:2009zh}. In this paper we
examine the simplest GUT example where this phenomenon is closely
related to the see-saw mechanism \cite{Minkowski} for generating the
neutrino masses and may provide a link between axions and right-handed neutrinos. \\
At the same time our scenario establishes a possible link between leptogenesis and dark matter  \cite{DiBari:2019amk,DiBari::2017iyk,DiBari:2017uka} in a generalized setting, due to the prediction of an axion in the low energy spectrum. 
Stueckelberg axions $(b(x))$ appear in the field theory realization of the Green-Schwarz mechanism of anomaly cancelation of string theory, in the dualization of a 3-form, and correspond to pseudoscalar gauge degrees of freedom (see also the discussion in \cite{Coriano:2018uip}). As ordinary Nambu-Goldstone modes they undergo a local shift 
\begin{equation}
b(x)\to b(x) + M \theta(x)
\end{equation}
under an abelian gauge transformation and are coupled to the anomaly via 
a dimension-5 operators of the form ${b(x)}/{M} F\wedge F$ where $F$ is, generically, the field strength of the  gauge fields which share a mixed anomaly with the $U(1)$ symmetry, and $M$ is the Stueckelberg scale. \\
In these scenarios, pseudoscalar gauge degrees of freedom may develop physical components only after the breaking of the shift symmetry by some extra potential. This is expected to occur in the case of phase transitions in a non-abelian gauge theory, when instanton interactions naturally arise and induce a mixing between the Stueckelberg field and the Higgs sector of the theory, with the generation of a periodic potential, after spontaneous symmetry breaking. \\ 
This scenario in which the CP odd phases of the scalar sector mix and generate such a potential, has provided the basic template for the emergence of a physical CP odd state, in a way which is very close to what was conjectured to occur in the case of the electroweak or DFSZ version of the Peccei-Quinn \cite{Peccei:1977ur} axion (see the review \cite{Kim:2008hd}), where the anomalous symmetry is a global rather than a local one. \\
Indeed, we recall that in the DFSZ case one writes down a general potential, function of three scalar fields, which is $SU(2)\times U(1)$ invariant. The simplest realization of this scenario is in the two-Higgs doublet model, where the Higgs fields $H_u$ and $H_d$ are assigned the global symmetry 
\beq
H_u\to e^{i\alpha X_u }H_u,\qquad  H_d\to e^{i\alpha X_d }H_d \qquad
\eeq
under $U(1)_{PQ}$
and are accompanied by an additional scalar $\Phi$, which is singlet under the Standard Model (SM) symmetry
\beq
 \Phi\to e^{i\alpha X_\Phi}\Phi
\eeq
with  $X_u +X_d=-2 X_\Phi$. The potential is given by a combination of terms of the form
\beq
\label{mixing}
V= V( \vline H_u\vline^2 \,,\, \vline H_d\vline^2\, ,\,\vline \Phi\vline^2 \,,\,\vline H_u H_d^\dagger \vline^2\, , \,\vline H_u\cdot H_d \vline^2 \, ,\, H_u\cdot H_d, \Phi^2 ) 
\eeq
(with $H_u\cdot H_d\equiv H_u^\alpha H_d^\beta\epsilon_{\alpha\beta}$) which is invariant under the Standard Model gauge symmetry and is in addition invariant under the global $U(1)_{PQ}$. \\
As pointed out in \cite{Coriano:2009zh} a similar effective theory can be obtained in the case of a gauge symmetry, in a scenario that leaves most of the intermediate steps in the generation of Stueckelberg-like Lagrangian unchanged. In this realization of the Stueckelberg Lagrangian, the Stueckelberg pseudoscalar emerges from the phase of the complex scalar field which is responsible for the breaking of the gauged $U(1)$ symmetry. The breaking takes place at the GUT (Grand Unified Theory) scale, which takes the role of the Stueckelberg mass for the low energy effective theory. \\
In our case such abelian symmetry is contained within $SO(10)$ and it is identified with $U(1)_{X}$.  This provides the basic observation which motivates our work, which connects the decoupling of a gauge boson corresponding to an  $U(1)_{X}$ symmetry within $SO(10)$ and of a right-handed neutrino to the appearance of an axion in the spectrum of the low energy theory. Being the construction sequential in each of the three generations, this scenario predicts three axions in the spectrum. Building on a similar analysis by two of us in \cite{Coriano1} based on a $E_6\times U(1)_X$, such axions are expected to be ultralight, in the $10^{-20}$ eV 
mass range.

\subsection{Incomplete decoupling of a chiral fermion and global anomalous $U(1)_{X}$} 
We believe it is useful to scrutinise this within a transparent model
where two examples of physics beyond the standard model,
the non-zero neutrino masses and the Stueckelberg axion are
closely related. Since we know from experiment \cite{SuperK}
that the first extension exists in Nature, it increases our expectation
that the second should be realised. We shall review the group theory of SO(10) including the available
irreducible representations for the matter particles and the symmetry
breaking.\\
 SO(10) naturally provides three right-handed neutrinos
which can participate in the see-saw. Because of the decoupling
of these additional neutrino states at high masses, the resultant
effective theory possesses an anomalous $U(1)_{X}$
symmetry.
Since we shall discuss neutrino masses it is worth recalling the
various possibilities for introducing them into the minimal SM. We shall mention four of these, one being the
see-saw mechanism, and reveal why the other three are
less attractive. One of them, introduced in \cite{Zee},
once appeared to be compelling when based only
on the SuperKamiokande experiment \cite{SuperK} but it
predicted maximal solar neutrino mixing which unfortunately
was subsequently excluded by the SNO experiment \cite{SNO}.
This left as the most popular possibility the see-saw mechanism
which we shall employ in the present model. When neutrino masses were established
experimentally in 1998 there was confusion about to whom
priority for the see-saw idea belonged and it was temporarily assigned to a
number of theory 
papers published in 1979. Further scholarship revealed, however, that priority
belonged to a 1977 paper by Minkowski \cite{Minkowski}.

\section{SO(10) Grand Unification}

The SO(10) model for unifying quarks and leptons was invented over forty years ago
in \cite{Georgi,FM}. After non-zero masses
for neutrinos were discovered, it became the most popular
GUT superseding the otherwise more economical SU(5) GUT \cite{GG}.
A recent discussion of an SO(10) GUT can be found in \cite{Hall}.
In the minimal Standard Model (SM), as in the minimal SU(5) GUT, the neutrinos
were assumed to be massless. In the SO(10) GUT, each family in a {\bf 16} contains,
in addition to the fifteen helicity states of the minimal SM, a right-
handed neutrino $N$. This gives rise to several additional features, beyond the
most obvious one that the neutrinos can acquire mass through the see-saw
mechanism. 
An SU(5) GUT subsumes the SM gauge group $SU(3)_C \times SU(2)_L \times U(1)_Y$
but an SO(10) GUT with one additional rank includes also a $U(1)_{(X)}$. It
is this gauged $(X)$ symmetry and its breaking which will play a central
role in our present discussion.\\
The group theory underlying the SO(10) GUT is well-known and reviewed in many
papers; one reliable such reference is \cite{Slansky}. \\
For the purposes of establishing
notation we shall briefly discuss this with special emphasis on the role of (X)
symmetry which will be treated further in subsequent subsections.
\subsection{Breaking patterns}
The gauge group $SO(10)$ has the dimension 45 of its adjoint. An adjoint of scalars can
break the symmetry while preserving rank-5 to

\begin{equation}
SO(10) \rightarrow [SU(3)_C \times SU(2)_L \times U(1)_Y]_{SM} \times U(1)_{X}
\label{45breaking}
\end{equation}
with $3X = 12 Y - 15 (B-L)$.
We shall need more scalars to give mass to the fermions. Each family is in a {\bf 16}
irreducible representation. For example the first family is

\begin{equation}
16 \equiv (u^r, u^g, u^b, d^r, d^g, d^b; 
u^r, u^g, u^b, d^r, d^g, d^b; \nu_e, e^-, N, e^+)_L
\label{16}
\end{equation}
where we have designated the colours as $r, g, b$ (= red, green, blue).
The Yukawa couplings which can provide fermion masses require scalar fields which 
are included in 

\begin{equation}
16 \times 16 = 10_s + 120_a + 126_s
\label{16x16}
\end{equation}
where the subscripts $s, a$ specify symmetric, antisymmetric. The 10 is the vector representation
of SO(10), although the spinor representation 16 is really the defining representation, because one can
make 10 from 16, as in Eq.(\ref{16x16}), but not {\it vice versa}.
We first consider the decomposition of SU(5) into $SU(3)_c \times SU(2)_L \times U(1)_Y$,
adopting the notation $( SU(3)_C, SU(2)_L )_Y$ with the result that

\begin{eqnarray}
\bar{5} &=& (\bar{3}, 1)_{+2/3} + (1, 2)_{-1}   \nonumber \\
10 & = & (3, 2)_{+1/3} + (\bar{3}, 1)_{-4/3} + (1, 1)_{+2} \nonumber \\
\bar{15} & = & (6, 1)_{-4/3} + (3, 2)_{+1/3} + (1, 3)_{+2}     \nonumber \\
24 & = &  (8, 1)_0 + (3, 2)_{-5/3} + (\bar{3}, 2)_{+5/3}  + (1, 3)_0 + (1, 1)_0     \nonumber \\
45 & = &  (8, 2)_{+1} + (\bar{6}, 1)_{-2/3} + (\bar{3}, 2)_{-7/3} + (\bar{3}, 1)_{-4/3} +      \nonumber \\
&  & (3, 3)_{-2/3} + (3, 1)_{-2/3} + (1, 2)_{+1}     \nonumber \\ 
\bar{50} & = &  (8, 2)_{+1} + (6, 1)_{+8/3} + (\bar{6}. 3)_{-2/3} +  (\bar{3}, 2)_{-7/3} + \nonumber \\
& & (3, 1)_{-2/3} + (1, 1)_{-4}   
\label{SU5to321}
\end{eqnarray}
The states in the first two lines of Eq.(\ref{SU5to321}) are the familiar ones
of one SM family, without a right-handed neutrino, which is why $(10+\bar{5})$
is used in an SU(5) GUT. The scalars in the SU(5) Yukawa couplings
must be among
\begin{eqnarray}
\bar{5} \times \bar{5} = 10_a + 15_s   \nonumber \\
10 \times \bar{5} = 5 + 45  \nonumber \\
10 \times 10 =   \bar{5}_s + \bar{45}_a + \bar{50}_s  
\label{SU5Yukawas}
\end{eqnarray}
and we note that the usual Higgs boson, which in this notation is the complex doublet
$(1, 2)_{\pm 1}$, appears uniquely in the {\bf 5} and {\bf 45} of SU(5), as can be seen
from Eq.(\ref{SU5to321}).
Armed with these preliminaries about SU(5), it is rendered almost trivial to extend the
analysis to SO(10), but the (X) symmetry means we must tread carefully.
We return to Eq.(\ref{16x16}) and adopt a new notation in the SO(10) decompositions
of $(SU(5))_{X}$. From \cite{Slansky} we are able to decompose the scalar
SO(10) irreducible representations into their SU(5) components:

\begin{eqnarray}
10 & = & 5_2 + \bar{5}_{-2}     \nonumber  \\
120 & = & 5_2 + \bar{5}_{-2} + 10_{-6} + \bar{10}_6 +45_2 +\bar{45}_{-2}  \nonumber \\
126 & = & 1_{-10} + \bar{5}_{-2} +10_{-6} + \bar{15}_6 +45_2 + \bar{45}_{-2} \nonumber \\
45 &=& 24_0 + 10_4 + \bar{10}_{-4} +1_0 
\label{SO10toSU5U1}
\end{eqnarray}
All of $10$, $120$ and $126$ necessarily contain a candidate for the SM complex Higgs doublet.
From Eq. (\ref{SU5to321}), we can, if needed, translate the SU(5) representations
in Eq.(\ref{SO10toSU5U1}) into SM representations. This provides all the group
theory we shall need in the present article.
In the following we shall focus on the breaking of $U(1)_{(X)}$ which is intimately
related to the mass of the right-handed neutrinos N in Eq.(\ref{16})
and hence to the see-saw mechanism. \\
\subsection{The two complex singlet scalars in the effective potential}
If we introduce a
scalar field $\Phi$, singlet under SU(5) with lepton number L=+2, we can write
the Majorana mass M of the right-handed neutrino $N_R^i$ (i,j =1,2,3) of the three generations as

\begin{equation}
\lambda_{i j} N_R^i N_R^j \Phi .
\label{Majorana}
\end{equation}
The masses $\lambda_{ij}\langle \Phi\rangle$  may be taken to be $\sim 10^{10}$ GeV, far above the weak scale,
whereupon we may integrate out the right-handed neutrino N to derive an effective field 
theory with interesting properties. In particular, the gauged $U(1)_{(X)}$
of the SO(10) GUT has become anomalous, because in the $(X)^3$ triangle
diagram N has been removed from the internal states.\\
We note that the 126 of scalars in Eq.(\ref{SO10toSU5U1})
contains an SU(5) singlet, charged under $(B-L)$, in addition to the SU(5) singlet in the 45 of Eq. \ref{SO10toSU5U1}, $\Phi$. The presence
of two such states in our model will be relevant in our subsequent analysis.\\
Let us step back to a purely bottom-up approach. Consider
the original minimal standard model (MSM) with massless neutrinos. In perturbation
theory, it conserves baryon number (B) and lepton number (L) so there is a
global $U(1)_{(B-L}$ which, without a right-handed neutrino, is anomalous.
Such a statement is obviously not connected to grand unification.
Of course, this model is ruled out because neutrinos have non-zero masses
so some modification is necessary to the MSM and there is a number
of possibilities\cite{FG}. The most popular is the addition of
right-handed neutrinos which permit the see-saw mechanism for generating
neutrino masses. This is achieved most naturally in $SO(10)$ unification.

Now we carefully discuss a top-down analysis of $SO(10)$ spontaneous symmetry
breaking. At the GUT scale ($10^{15-16}$ GeV) the adjoint 45 is used to
break the symmetry in a necessarily rank-preserving manner according to
\begin{equation}
SO(10) \rightarrow SU(5) \times U(1)_{X} \rightarrow SU(3) \times SU(2) \times U(1)_{Y} \times U(1)_X
\end{equation}
so that the $U(1)_{X}$, with $3X=12Y-15(B-L)$, is still unbroken and its gauge boson is massless.
At an intermediate scale $M_I \sim 10^{10-11}$ GeV the complex 126 is
used spontaneously to break $U(1)_{X}$ and to give Majorana masses to
the three right-handed neutrinos. This arises from a VEV of the $SU(5)$-singlet
complex component in Eq.(\ref{126}) which has the Mexican-hat type of potential
required for the Higgs mechanism.

\section{See-Saw Mechanism}
In the MSM neutrinos are massless. 
The minimal standard model involves three chiral neutrino states, 
but it does not admit renormalizable interactions that can generate 
neutrino masses. Nevertheless, experimental evidence suggests 
that both solar and atmospheric neutrinos display flavor oscillations, 
and hence that neutrinos do have mass. Two very different neutrino 
squared-mass differences are required to fit the data:

\begin{equation}
6.9 \times 10^{-5} {\textrm eV^2} \leq \Delta_s \leq 7.9 \times 10^{-5} {\textrm eV^2}  ~~ {\rm and} ~~ \Delta_a \sim (2.4 - 2.7) \times 10^{-3} {\textrm eV^2}, 
\label{massdiff}
\end{equation}
where the neutrino masses $m_i$ are ordered such that:

\begin{equation}
\Delta_s = |m_2^2 - m_1^2| ~~ {\rm  and} ~~  \Delta_a = |m_3^2 - m_2^2| \simeq |m_3^2 - m_1^2|
\label{diffmass}
\end{equation}
and the subscripts s and a pertain to solar (s) and atmospheric (a) oscillations respectively. The large uncertainty in $\Delta_s$ reflects the several potential explanations of the observed solar neutrino flux: in terms of vacuum oscillations or large-angle or small-angle MSW solutions, but in every case the two independent squared-mass differences must be widely spaced with

\begin{equation}
r= \Delta_s/\Delta_a \sim 3\times 10^{-2}.
\label{r}
\end{equation}
In a three-family scenario, four neutrino mixing parameters suffice to describe neutrino oscillations, akin to the four Kobayashi-Maskawa parameters in the quark sector. Solar neutrinos may exhibit an energy-independent time-averaged suppression due to $\Delta_a$, as well as energy-dependent oscillations depending on $\Delta_s/E$. Atmospheric neutrinos may exhibit oscillations due to $\Delta_a$, but they are almost entirely unaffected by 
$\Delta_s$. It is convenient to define neutrino mixing angles as follows:

\begin{equation}
\left( \begin{array}{c} \nu_e \\ \nu_{\mu} \\\nu_{\tau}
\end{array} \right)
=
\left( \begin{array}{ccc} c_2c_3  &  c_2s_3  & s_2 e^{-i\delta} \\
+ c_1s_3 + s_1s_2c_3e^{i\delta} & -c_1c_3 - s_1s_2s_3e^{i\delta} & -s_1c_2 \\
+s_1s_3 - c_1s_2 c_3 e^{i\delta} &  -s_1c_3 - c_1s_2s_3e^{i\delta} & +c_1c_2
\end{array} \right)
\left( \begin{array}{c} \nu_1 \\ \nu_2 \\\nu_3
\end{array} \right)
\label{PMNS}
\end{equation}
with $s_i$ and $c_i$ standing for sines and cosines of $\theta_i$. 
For neutrino masses satisfying (\ref{massdiff}), the
vacuum survival probability of solar neutrinos is:

\begin{equation}
P (\nu_e \rightarrow \nu_e)|_s \simeq 1 - \frac{\sin^2 2\theta_2}{2} - \cos^4 \theta_2 \sin^2 2\theta_3 \sin^2
\left( \Delta_s R_s/4E \right)
\label{nuenue}
\end{equation}
whereas the transition probabilities of atmospheric neutrinos are:
\begin{eqnarray}
P(\nu_{\mu} \rightarrow \nu_{\tau})|_a & \simeq & \sin^2 2\theta_1 \cos^4\theta_2  \sin^2 
\left( \Delta_a R_a/4E \right)  \nonumber \\
P(\nu_e \rightarrow \nu_{\mu})|_a & \simeq & \sin^2 2\theta_2 \sin^2\theta_1  \sin^2 
\left( \Delta_a R_a/4E \right)  \nonumber \\
P(\nu_e \rightarrow \nu_{\tau})|_a & \simeq & \sin^2 2\theta_2 \cos^2\theta_1 \sin^2 
\left( \Delta_a R_a/4E \right) 
\label{transition}
\end{eqnarray}
None of these probabilities depend on $\delta$, the measure of CP violation.
Let us turn to the origin of neutrino masses. Among the many renormalizable and
gauge-invariant extensions of the standard model that can do the trick are \cite{FG}
(i) The introduction of a complex triplet of mesons $(T^{++}, T^+, T^0)$ coupled bilinearly to pairs of lepton doublets. They must also couple bilinearly to the Higgs doublet(s) so as to avoid spontaneous $(X)$ violation and the appearance of a massless and experimentally excluded majoron. This mechanism can generate an arbitrary complex symmetric Majorana mass matrix for neutrinos.
(ii) The introduction of singlet counterparts to the neutrinos with very large Majorana
masses. The interplay between these mass terms and those generated by the Higgs boson, the so-called see-saw mechanism, yields an arbitrary but naturally small Majorana neutrino mass matrix.
(iii) The introduction of a charged singlet meson $f^+$ coupled antisymmetrically to pairs of lepton doublets, and a doubly-charged singlet meson $g^{++}$ coupled bilinearly both to pairs of lepton singlets and to pairs of f-mesons. An arbitrary Majorana neutrino mass matrix is generated in two loops.
(iv) The introduction of a charged singlet meson $f^+$ coupled antisymmetrically to pairs of lepton doublets and (also antisymmetrically) to a pair of Higgs doublets. This simple mechanism was first proposed in \cite{Zee}  and results at one loop in a Majorana mass matrix in the flavor basis $(e, \mu, \tau)$ of a special form:
\begin{equation}
\left( \begin{array}{ccc}
0 & m_{e\mu} & m_{e\tau} \\
m_{e\mu} & 0 &m_{\mu\tau} \\ 
m_{e\tau} & m_{\mu\tau} & 0
\end{array} \right)
\label{zee}
\end{equation}

\bigskip

\noindent
This Zee model is attractive as an simple extension of the SM. It predicts maximal
solar neutrino mixing, $\theta_{12} = \frac{\pi}{4}$, a value which was strongly disfavoured
by SNO data\cite{SNO,FOY}. Of all the models preserving only the three chiral left-handed
neutrinos of the SM - models (i), (iii) and (iv) above - model (iv) is surely the most
appealing and it fails. Therefore one is led to additional neutrino states, typically
two or more massive right-handed neutrinos which we denote $N_I$ ($i=1,2,\ldots,p$).\\
In the model we shall discuss $p$ is necessarily $p=3$ because each of the three quark-lepton
families is in a {\bf 16} of $SO(10)$ and each contains one $N$ state. There has been
considerable interest in more minimal models with $p=2$ as introduced in the so-called
FGY model of \cite{FGY}. This choice has the property of reducing the number of free
parameters such that the CP-violating phase in $N_i$ mixing matrix is simply related
to the CP-violating phase, $\delta$, in Eq.(\ref{PMNS}). This means that the measurement
of $\delta$ in long-baseline neutrino oscillation experiment would shine light on the
origin of matter-antimatter asymmetry arising from leptogenesis\cite{FY} where it
arises from $N_i$ decay. In general, this connection does not exist so that an
optimistic logic could argue that the FGY model, sometimes called the minimal see-saw,
is possibly correct.\\
For the present case of $p=3$ we introduce a mass basis

\begin{equation}
(\nu_e. \nu_{\mu}. \nu_{\tau}, N_1, N_2, N_3)
\label{basis}
\end{equation}
so that there is a $6 \times 6$ mass matrix in four $3 \times 3$ blocks with
the top-left block vanishing and the bottom-right being the large Majorana
masses for the $N_i$. The two off-diagonal blocks are Dirac masses coupling
the $\nu_{IL}$ to the $N_{iR}$.

The effective mass matrix of the light Majorana neutrinos is given by
\begin{equation}
M =  M_D (M_R)^{-1} M_D^T 
\label{seesaw}
\end{equation}
\noindent
where $M_D$ and $M_R$ are the $3 \times 3$ mass matrices for the Dirac and right-handed
Majorana neutrinos,
respectively. $M_D^T$ designates the transpose. \\
The see-saw strategy is immediately evident from Eq.(\ref{seesaw}). Denoting
the mean values of the $3\times3$ blocks by $m$ and $M$

\begin{equation}
\left( \begin{array}{cc} 0 & m \\
m & M 
\end{array}
\right)
\label{blocks}
\end{equation}
the eigenvalues for $m \ll M$ are close to $m^2/M$ and $M$. This shows how large
the $N_i$ masses are expected to be. Taking the first family, with a typical quark
mass $10$ MeV and electron neutrino mass $10^{-5} eV$, we find $M\sim 10^{10}$ GeV.
Coincidentally, and suggestively, such a mass fits well with the mass required 
for successful leptogenesis\cite{FY}.\\
This discussion exhibits the great advantage of the see-saw mechanism compared to
the alternative models discussed above: the smallness of the neutrino masses relative
to those of the quarks and leptons occurs naturally. That being said, the other side of
the coin is that experimental observation of the very massive $N_i$ is challenging.\\
The crucial observation for our present purposes is to consider the $U(1)_{X}$
triangle anomalies. If we keep all the states in Eq.(\ref{16}) for one family
\begin{equation}
16 \equiv (u^r, u^g, u^b, d^r, d^g, d^b; 
u^r, u^g, u^b, d^r, d^g, d^b; \nu_e, e^-, N, e^+)_L,
\label{162}
\end{equation}
then we can examine this question.\\
The pure gauge anomaly $U(1)_{X}^3$ has cancelling contributions from the states in Eq.(\ref{162})
as follows

\begin{equation}
6 \left( \frac{1}{27} \right) + 6 \left( - \frac{1}{27} \right) +2 (+1) +2(-1) = 0
\label{B-Lcube}
\end{equation}
For the gravitational triangle anomaly which has only one $U(1)_{X}$ vertex
the respective cancelling contributions are
\begin{equation}
6 \left( \frac{1}{3} \right) + 6 \left( - \frac{1}{3} \right) +2 (+1) +2(-1) = 0.
\label{B-L}
\end{equation}
When we decouple the $N$ state in Eq.(\ref{162}) by taking it to very
high mass, the right hand sides of Eq.(\ref{B-Lcube})and Eq.(\ref{B-L}) both change from
zero to $-1$, the anomalies do not cancel, and therefore there exists in the effective
theory an anomalous $U(1)$ symmetry of the sort considered
in different contexts in {\it e.g.} \cite{U1A1,U1A2,U1A3,U1A4}.

\section{Anomalous $U(1)_{X}$}
Let us introduce the matter fields in our model. The fermions
are in three {\bf 16}'s, $\Psi_i$ ($i=1,2,3$). Each {\bf 16} contains
a right-handed neutrino $N_R^i$ with $(X) = +1$.\\
$SO(10)$ contains the usual $SU(5)$ subgroup \cite{GG}
which plays a r\^{o}le in containing the minimal standard
model (MSM) as if without neutrino mass. To provide mass
to $N_R$ without breaking $SU(5)$ we introduce a
complex scalar $\Phi$ in the {\bf 126} of $SO(10)$
which under $SU(5)$ contains

\begin{equation}
126 \subset 1 + 5 + \bar{10} + 15 + \bar{45} + 50
\label{126}
\end{equation}
and the $N_R^i$ acquire mass as in Eq. \ref{Majorana}
when the $SU(5)$-singlet component of $\Phi$
in Eq.(\ref{126}) gains an intermediate mass scale VEV
\begin{equation}
< \Phi > = M_I
\label{MI}
\end{equation}
where for the see-saw mechanism the intermediate
mass scale $M_I$ is typically $\sim 10^{10}$ GeV.\\
To break the symmetry $SU(5)$ to that of the standard
model we introduce more scalars in the representations
of $SO(10)$ which are the adjoint $A$ in a {\bf 45},
the vector $V$ in a {\bf 10} and finally a spinor $B(16)$.\\
The adjoint {\bf 45} decomposes under $SU(5)$ as
\begin{equation}
45 \supset 1 + 10 + \bar{10} + 24
\label{45}
\end{equation}
so that the {\bf 24} can provide the rank-preserving
$SU(5) \rightarrow SU(3) \times SU(2) \times U(1)$.\\
We recall that in $SO(10)$, 45 decomposes as in Eq. \ref{SO10toSU5U1} within which the 24 can provide the rank-preserving $SU(5)\to SU(3)\times SU(2)\times U(1)_Y$ symmetry breaking.
The fermion masses arise from the Yukawa couplings
\begin{equation}
{\cal L}_{Yukawa} = \Psi \left(Y_V V + Y_{\Phi} \Phi \right) \Psi
\label{Yukawa}
\end{equation}
which may be understood to contain the coupling of Eq.(\ref{Majorana}).\\
We adopt the convention that Latin indices $a,b,c,\ldots$ run from
$1$ to $10$ and Greek indices $\alpha,\beta,\gamma,\ldots$
run from $1$ to $16$. The vector field $V$ is $V_a$ and the
adjoint $A$ is $A_{ab}=-A_{ba}$ so that all the $V$ and $A$
couplings up to quartic in the Higgs potential can be written,
bearing in mind that
\begin{eqnarray}
10 \times 10 &\supset& 1 + 45 + 50  \nonumber \\
10 \times 45 &\supset& 10 + 120 +320 \nonumber \\
45 \times 45 &\supset& 1 + 45 + 54 + 210 + 770 + 945. 
\label{1045}
\end{eqnarray}
in the form
\begin{equation}
{\cal V}(V,A) = V_aV_a + (V_aV_a)^2 + A_{ab}A_{ab}  +(A_{ab}A_{ab})^2
+ (V_aV_a)(A_{bc}A_{bc}) +\ldots
\end{equation}
among other terms.\\
To deal with the {\bf 126} it is essential to introduce the $\Gamma$ matrices
\begin{equation}
\Gamma^a_{\alpha\beta}
\end{equation}
which are ten $16 \times 16$ matrices which roughly generalise the four $4 \times 4$
Dirac matrices $\gamma^{\mu}$ pertinent to $O(4)$, and likewise satisfy a Clifford algebra.
The $\Phi$ field of the {\bf 126}
is a symmetric scalar field satisfying the trace condition
\begin{equation}
\Gamma^a_{ij}\Phi_{ji} = Tr (\Gamma^a \Phi) = 0
\label{trace}
\end{equation}
Now, in addition to Eq.(\ref{1045}), we shall need
\begin{eqnarray}
126 \times 10 & \supset & 210 + 1050.   \nonumber \\
126 \times 45 & \supset & 120 +126 + 1728 + 3696. \nonumber \\
126 \times 126 & \supset & 54_S + 945_A + 1050_S +2772_S + 4125_S + 6930_A.
\label{126}
\end{eqnarray}
to write the Higgs potential terms involving $\Phi$ such as
\begin{eqnarray}
{\cal V}(\Phi) & = & \Phi_{ij}\Phi_{ij} +  (\Phi_{ij}\Phi_{ij})^2 + \Phi_{ij}\Phi_{jk}\Phi_{kl}\Phi_{li} \nonumber \\
        & &  + \Gamma^a_{ij} \Phi_{jk}\Phi_{kl}\Phi_{lm}\Phi_{mn}\Gamma^a_{ni} + \ldots
        \label{Vphi}
        \end{eqnarray}
among other terms including mixed $\Phi-A$ terms possible under $SO(10)$ symmetry, as can be
seen from Eqs.(\ref{1045}) and (\ref{126}). We take note of the cubic scalar coupling
$16.16.\overline{126}$ which may be written
\begin{equation}
B_{\alpha}B_{\beta}\Phi^*_{\alpha\beta}
\label{BBPhi}
\end{equation}
and which we shall use in the next section.

\section{Stueckelberg Axion}

In order to illustrate how the mixing of the CP-odd phases takes place in the breaking of $SO(10)\to SU(5)\times U(1)$ we consider specific terms in the potential, describing the conditions which need to be satisfied in order to generate a periodic potential function of a single gauge invariant field. The latter takes the role of a physical axion and will be denoted by $\chi$. \\
The periodic potential is generated at the scale at which $SU(5)\times U(1)$ is broken necessarily at
$ > 10^{15}$ GeV to avoid too-fast proton decay. At this GUT scale, instanton effects are present. In order to 
understand why this happens, we consider an $SO(10)$ invariant term in the original theory such as 
\begin{equation} 
 16\times 16\times \overline{126}
\end{equation} 
which is 
built out of the spinorial (16) of $SO(10)$ and the complex conjugate of the 126. The $SO(10)$ singlet is obtained from
\begin{equation} 
16\times 16=10_s + 120_a + 126_s 
\end{equation} 
by combining the $126_s$ taken from the symmetric part of the product $(16\times 16)_s=126_s + 10_s$ with the $\overline{126}$.
We can specialize \eqref{16x16} by indicating the $X$ content of the decomposition 
using 
\begin{equation}
16=1_{-5} +\bar{5}_{+3} +10_{-1}
\end{equation}
from which gives for their antisymmetric product
\begin{eqnarray}
120_a&=&(16\times 16)_a\nonumber \\
&=&\bar{5}_{-2} +10_{-6}+ (5 + 45)_{+2} + \bar{10}_{+6} +\overline{45}_{-2}
\end{eqnarray}
while the symmetric component can be specialized in the form 
\begin{eqnarray}
(16\times 16)_s &=&126_s + 10_s \nn\\
&=& (1_{-10} +\bar{5}_{-2} +10_{-6} +\overline{15}_{+6} + 45_{+2} +\overline{50}_{-2}) + 
(5_{+2} +\bar{5}_{-2})
\label{first}
\end{eqnarray}
where the two contributions in brackets refer respectively to the $126_s$ and to the $10_s$ of $SO(10)$.\\
A periodic potential can be extracted from the decomposition above starting from 
the $126_s\times \overline{126}$, $SO(10) $ singlet, by combining the $1_{-10}$ in Eq. \eqref{first} with the $1_{+10}$ in the $\overline{126}$, the latter obtained by conjugation of \eqref{126} - with the inclusion of its complete $SU(5)\times U(1)_{X}$ content - 
\begin{equation}
\overline{126}=1_{+10} +5_{+2} +\overline{10}_{+6}+15_{-6}+\overline{45}_{-2} +50_{+2}.
\label{1262}
\end{equation}
A term in this form in the potential allows to induce a mixing of the CP-odd phases of the two SU(5) singlet representations in such a way that one linear combination of these will correspond to a physical axion while the second one will be part of the Nambu-Goldstone mode generated by the breaking of $U(1)_{X}$.\\
We will be denoting with $\sigma$ and $\phi$ the two fields corresponding to the $1_{-10}$ and $1_{10}$ respectively, denoting their vevs with $v_\sigma$ and $v_\phi$ respectively. We will assume that $v_\phi$ will be large in such a way to provide a mass term for the right-handed neutrino, as specified in \eqref{Majorana} using the Majorana operator $N_R N_R \phi$.

In order to characterise the structure of the Stueckelberg Lagrangian at classical level we focus our attention on the extra (periodic) potential related to $\sigma$ and $\phi$
\begin {equation}
V_p=\lambda M_I^2 \sigma\phi + \textrm{ h.c.}
\end{equation}
Since there must be an $SU(5)$ singlet it is important to realise that the other parts of Eq.(\ref{1262}) do not contribute.
The coupling $\lambda$ is instanton generated at the scale $M_{GUT}$, a fact which provides a drastic suppression in $V_p$. We parameterize both fields around their vevs as
\begin{eqnarray}
\sigma&=&\frac{v_\sigma + \sigma_1 + i \sigma_2}{\sqrt{2}}\nonumber\\
      &=& \frac{v_\sigma +\rho_\sigma}{\sqrt{2}}e^{i F_\sigma(x)/(g_B v_\sigma)} \nonumber \\
\phi&=&\frac{v_\phi + \rho_\phi}{\sqrt{2}}e^{i b(x)/ v_\phi} \nonumber \\
\end{eqnarray}
and $v_\phi$ is at the GUT scale $M_{GUT}\sim 10^{15}$ GeV. The parameterization of $V_p$ in a broken phase is made possible by the remaining 
- non periodic - general scalar potential which will assume a typical mexican-hat shape as for an ordinary $U(1)$ symmetry. Both $\sigma$ and $\phi$ are charged under $U(1)_{(X)}$ and therefore their vevs break the gauged  $(X)$ which as we have discussed survives as an anomalous $U(1)$
in the effective theory at low energies.  
We denote with $g_B$ the gauge coupling of the $U(1)_{X}$ gauge boson $(B_\mu)$, while $\pm q_B$ will denote the corresponding $X$ charges of the scalars. 
Their normalization, equal to $\pm 10$ in the normalization of \cite{Slansky}, is indeed arbitrary. The role of the Stueckelberg field is taken by $b(x)$ in the polar parameterization of $\phi$, which is normalized to 1 in mass dimension, while $F_\sigma$ is massless.

The two covariant derivatives of the scalars take the form 
\begin{eqnarray}
D_\mu \sigma&=&\left(\partial_\mu + i q_B g_B\, B_\mu\right) \sigma\nonumber \\
D_\mu \phi&=&\left(\partial_\mu  +  i q_B g_B \, B_\mu\right) \phi
\end{eqnarray}
with the typical Stueckelberg kinetic term generated from the decoupling of the radial fluctuations of the $\phi$ field
\begin {equation}
| D_\mu \phi |^{\,\,2}=\frac{1}{2}\partial_\mu\rho_\phi\partial^\mu\rho_\phi + \frac{1}{{2}}
( \partial_\mu b - M B_\mu)^2
\end{equation}
with $M=q_B g_B v_\phi\sim M_I$ takes the role of the Stueckelberg scale. In general it is natural to assume that both $v_\phi$ and $v_\sigma$ are of the same order, and the mass of $B_\mu$, the X gauge boson, will be given as a mean of both vevs
\begin{equation}
M_B=\sqrt{ (q_B g_B v_\sigma)^2 + M^2} 
\end{equation}
The quadratic action, neglecting the contribution of the radial excitations of $\sigma$ and $\phi $, can be easily written down for such $\sigma-\phi$ combination 
\begin{eqnarray}
\mathcal{L}_q &=&
\frac{1}{2} \left(\partial_\mu \sigma_{2}\right)^2 + \frac{1}{2}\left(\partial_\mu b\right)^2  
+ \frac{1}{2} M_B^2 B_\mu B^{\mu} 
  \nonumber\\
&& + B_\mu \partial^\mu \left(M_1  b +  v_\sigma g^{}_{B} q^{}_{B} \sigma_{2} \right),   
\end{eqnarray}
from which, after diagonalization of the mass terms we obtain 

\beqa
\mathcal{L}_q &=&\frac{1}{2} \left(\partial_\mu \chi^{}_{B}\right)^2 +
\frac{1}{2} \left(\partial_\mu G^{}_{B}\right)^2  + \frac{1}{2}\left(\partial_\mu h^{}_{1} \right)^2  
 + \frac{1}{2} M_B^2 B_\mu B^{\mu} - \frac{1}{2} m_{1}^2 h_{1}^2     \nonumber\\
&& + M_B B^\mu\partial_\mu G^{}_{B}. 
\eeqa
where we are neglecting all the other terms generated from the decomposition which will not contribute to the breaking. 
We can identify the linear combinations 

\beqa
\chi^{}_{B} &=& \frac{1}{M_B} \left(- M^{} \, \sigma^{}_{2} + q^{}_{B} g^{}_{B} v_\sigma \, b\right),   \nonumber\\
G^{}_{B} &=& \frac{1}{M_B}\left(q^{}_{B} g^{}_{B} v_\sigma \, \sigma^{}_{2} + M \, b\right), 
\eeqa
corresponding to the physical axion $\chi^{}_{B}$, and to a massless Nambu-Goldstone mode $G^{}_{B}$. 
The rotation matrix that allows the change of variables $(\sigma^{}_{2},b) \to (\chi,G^{}_{B})$ is given by 

\beq
U=\left(
\begin{array}{ll}
 -\cos \theta^{}_{B} & \sin \theta^{}_{B} \\
 \sin  \theta^{}_{B} & \cos \theta^{}_{B}
\end{array}
\right)
\eeq

with 
\begin{equation}
\theta^{}_{B}={\arcsin}(q^{}_{B} g^{}_{B} v_\sigma/ M_B). 
\end{equation}
The potential, as shown in similar analysis \cite{Coriano2}, is periodic in $\chi/f_\chi$
where $f_\chi\sim M_I$ takes the role of the axion decay constant. 
As already stressed before, the origin of this potential is nonperturbative and linked to the presence of instantons at the SO(10) GUT phase transition. For such reason, the size of the constants $\lambda$ in such potential are exponentially suppressed with $\lambda_i\sim e^{-2 \pi /\alpha_{GUT}}$, with the value of the coupling $\alpha_{GUT}$ fixed at the scale $M_{GUT}$ when the SO(10)
instantons are exact. The value of $\alpha_{GUT} $ here is in the range $1/33 \leq \alpha_{I} \leq 1/32$, 
giving $10^{-91} \leq \lambda_{i j} \leq 10^{-88}$, determining an axion mass given by 
$m_{\chi}^2 \sim  \lambda M_I^2$ in the range 
\begin{equation}
10^{-22} {\textrm eV} < m_{\chi} < 10^{-20}{\textrm eV}
\end{equation}
corresponding to an ultralight axion, which has been invoked for the resolution of several astrophysical constraints\cite{HOTWitten}.
\section{Conclusions}
We have investigated the possibility that the decoupling of a 
right-handed neutrino in the context of an $SO(10)$ GUT can be accompanied by an axion-like 
particle. Such a particle shares many of the properties already considered for a similar model discussed by two of us in the context of an $E_6\times U(1)_X$ unification, interpreted as low-energy GUT theory derived from string theory \cite{Coriano1}. 

\bigskip

\noindent
While, in the previous construction, the Stueckelberg Lagrangian was generated by the dualisation of a 3-form and required an anomalous $U(1)$ gauge symmetry, in this construction we have simply considered the possibility that the $U(1)_{X}$ symmetry of the Standard Model has an interesting implication. \\

\noindent
Starting from an $SO(10)$ symmetry, broken to an $SU(5)\times U(1)_{X}$ GUT symmetry, the decoupling of a right-handed neutrino leaves at low energy an action which is Stueckelberg like, with a global anomaly which couples to a CP-odd phase, $\chi(x)$. We have invoked the generation of a periodic potential in the $SU(5)\times U(1)_{X}$ effective theory in order to extract such gauge invariant degree of freedom in the 
pseudoscalar sector which couples to a global anomaly. 
Such Stueckelberg-like pseudoscalars are expected to be ultralight, around $10^{-{20}}$ eV  and to
decouple at the scale corresponding to the mass of the right-handed neutrino. An earlier paper which 
relates the lightness of the axion to neutrino mass is \cite{MS}. \\

\noindent
 We have illustrated, by analysing the representation content of the scalar sector of the $SO(10)$ and $SU(5)\times U(1)_{X}$ theories how this could be achieved. \\
 We believe that we have merely identified the general tracts of this mechanism to which we hope to return in the near future in a more extensive analysis.

\section*{Acknowledgements}
The work of C.C. is partially supported by INFN Iniziativa Specifica QFT-HEP.\\
 P.H.F. thanks INFN for financial support.

\end{document}